# Advancing Positron Emission Tomography Image Quantification Artificial Intelligence-Driven Methods, Clinical Challenges, and Emerging Opportunities in Long-Axial Field-of-View Positron Emission Tomography/Computed Tomography Imaging

Fereshteh Yousefirizi[1], Movindu Dassanayake[2], Alejandro Lopez[3], Andrew Reader[4], Gary J.R. Cook[4], Clemens Mingels[3,5], Arman Rahmim[1,6], Robert Seifert[3], Ian Alberts[6,7]

1. Department of Integrative Oncology, BC Cancer Research Institute, Vancouver, BC, Canada
2. School of Biomedical Engineering & Imaging Sciences, Kings College London, London, United Kingdom
3. Department of Nuclear Medicine, Inselspital, Bern University Hospital, University of Bern, Bern, Switzerland
4. King's College London and Guy's & St Thomas' PET Centre, School of Biomedical Engineering and Imaging Sciences, King's College London, London, United Kingdom
5. Department of Radiology, University of California Davis, Sacramento, CA, USA
6. Department of Radiology, University of British Columbia, Vancouver, BC, Canada
7. Molecular Imaging and Therapy, BC Cancer, Vancouver, BC, Canada

**Corresponding author:**
Ian Alberts, MA MBBS MD PhD
Molecular Imaging and Therapy
600 West 10th Ave
V5Z 4E6
Vancouver, British Columbia, Canada
ian.alberts@ubc.ca

**Clinical care points**

- MTV is increasingly recognized as an accurate estimate of disease burden and which has prognostic value, but its implementation has been hindered by the time-consuming need for manual segmentation of images. Automated quantitation using AI-driven approaches are promising.

- AI-driven automated quantification significantly reduces labor-intensive manual segmentation, improving consistency, reproducibility, and feasibility for routine clinical practice.
- AI-enhanced radiomics provides comprehensive characterization of tumor biology, capturing intratumoral and intertumoral heterogeneity beyond what conventional volumetric metrics alone offer, supporting improved patient stratification and therapy planning.
- AI-driven segmentation of normal organs improves radioligand therapy planning by enabling accurate dose predictions and comprehensive organ-based radiomics analysis, further refining personalized patient management.

# Introduction

Positron emission tomography/computed tomography (PET/CT) imaging plays a pivotal role in oncology, aiding tumor metabolism assessment, disease staging, and therapy response evaluation (1, 2). Traditionally, semi-quantitative metrics such as SUVmax have been extensively utilized, though these methods face limitations in reproducibility and predictive capability (3, 4). Recent advancements in artificial intelligence (AI), particularly deep learning, have revolutionized PET imaging, significantly enhancing image quantification accuracy, and biomarker extraction capabilities, thereby enabling more precise clinical decision-making (5-7).

The emergence of total-body (8) and long-axial field-of-view (LAFOV) (9) PET/CT scanners represents a transformative technological advance in PET imaging. Compared to standard axial field-of-view (SAFOV) scanners, LAFOV systems offer substantially extended axial coverage to simultaneously image larger body regions as well as increased sensitivity, up to an order of magnitude higher (10, 11), providing opportunities for enhanced clinical performance and novel research applications (12). These advancements allow for significant reductions in either injected radiotracer activity (13) or shortened scan times(14)(15, 16), and innovative imaging techniques such as multiplexed multi-tracer studies (17) and improved radiomics analyses (18). Notably, ultra-low dose imaging and robust normative databases facilitated by LAFOV PET/CT systems enable enhanced anomaly detection and broader clinical adoption, particularly when coupled with AI-driven methods for automated segmentation and quantification (19-23).

Although significant progress has been achieved, a clear clinical need persists for advanced, automated quantification techniques that can reliably extract clinically relevant biomarkers, such as total metabolic tumor volume (TMTV), from PET images (24-27). Manual segmentation, currently prevalent in clinical workflows, is labor-intensive and subject to high inter-observer variability, hindering the widespread clinical implementation of advanced PET quantification methods. AI-driven approaches, particularly deep learning-based segmentation and quantification techniques, promise to address these challenges effectively. These automated methodologies have demonstrated considerable potential across multiple malignancies, including lymphoma, prostate cancer, and neuroendocrine tumors, by providing accurate, reproducible, and scalable solutions that are increasingly critical in the era of LAFOV PET/CT (24-26).

To leverage these capabilities fully, robust infrastructures for data sharing, model training, and deployment, supported by interactive and adaptive AI tools, are essential. Collaborative, multi-center frameworks incorporating physician-in-the-loop systems, federated learning, and standardized data and model sharing protocols represent the future of scalable, clinically applicable PET quantification (28, 29).

## A clinical need for improved quantification? Identifying where advanced quantification and AI may have clinical impact

New techniques for enhanced image quantification, including improved image quality at the same or lower radiation exposure, and the extraction of new biomarkers for better patient stratification, are essential to advance the clinical application of nuclear medicine. When the utilization of tracers matures, leading to their use in routine clinical practice for (re)staging purposes, quantification-based research is conducted to leverage the full potential of the PET acquisitions. For example, PSMA-PET CT is the de facto reference standard for the staging and restaging of patients with prostate cancer (30-32). Disease progression can be assessed, either by measuring the tumor volume changes directly, or by frameworks like RECIP (33-37). Thereby, PSMA-PET is used to create a volumetric biomarker which correlates with disease extent (38), overall survival (39, 40) and prognosis (24, 35-37, 41).

The reduction of the three-dimensional PET acquisition to one solitary number like tumor volume might seem oversimplistic for an image-based phenotyping of the patient. For radioligand therapies, like PSMA-targeting [$^{177}$Lu]Lutetium PSMA-617, the degree of target expression and the disease extent are important to understand response pattern and select the best therapy candidates (42). The assessment of therapy eligibility is done by a simple visual analysis of the PSMA PET scan. However, new quantification methods are strongly associated with the response to therapy (41, 43) (**Figure 1**). Additionally [$^{18}$F]Fluorodeoxyglucose (FDG) may also be used for treatment selection, with preliminary evidence suggesting that it does not lead to a clinically significant management change compared with PSMA-PET assessment only (25, 44). The integration of FDG and PSMA PET derived biomarkers are still pending to date but could be helpful to better stratify patients for PSMA radioligand therapy, especially in earlier disease stages.

A second example of the clinical relevance of new quantification approaches is FDG-PET in lymphoma patients(45). Clinically, simplistic scores like the Lugano classification or the Deauville method are used to assess the disease extent or response to therapy (46-48). Even though they can in theory be collected by means of AI as recently shown for the Deauville score, they neglect a description of uptake heterogeneity, quantitative assessment of tumor bulks, tumor volume, etc (26). This is also relevant in the context of clinical trials, which use FDG-PET to decide on treatment escalation or de-escalation. For example, the PETAL trial that investigated PET-guided treatment intensification in aggressive non-Hodgkin lymphoma was negative and could not demonstrate a significant benefit of treatment intensification(27). An AI-based post hoc analysis that investigated the average FDG uptake in all lymphoma lesions could identify patients which benefitted from specific treatment arms(49). This highlights the need to integrate more complex quantification approaches in trial protocols, from which they may then translate into routine clinical practice. With the growing success of new tracers like fibroblast activation protein (FAP) inhibitors in gastrointestinal malignancies, a similar trend to PET-derived quantitative metrics will likely be seen in the future (50).

To implement new ways of quantification in PET data, manual solutions are often implemented which tend to be too labor intensive. This, along with a potentially low inter-reader repeatability, are two key reasons hindering the clinical adoption of advanced PET quantification approaches. To overcome both obstacles, AI methods are gaining increased attention. The transition from classification of threshold-segmented FDG-foci in either benign or malignant to automated U-Net-based segmentation of those regions in image space showcases the rapid development of such solutions (51-53). On a separate note, this now may also include multi-modal vision language models and large language models (LLMs), which could bridge the gap between the image domain and clinical information of referring physicians (54). As such, the potential benefit of LLMs might not be to write a textual PET report that faces the same limitations as a manual one but leverage existing text reports and PET data to come up with new biomarkers.

With the advent of digital and total body PET, the reduction of image noise from low dose examinations has been successfully introduced, even potentially enabling FDG- and PSMA-PET acquisitions as fast as CT scans (55, 56) (**figure 2**). In this context, algorithmic progress may facilitate the collection of novel quantitative PET features that have not been possible to extract before.

*Automated Disease Burden Quantification in PET/CT Images*

TMTV provides valuable insights into overall tumor burden, disease aggressiveness, and treatment response. Current clinical workflows for disease burden quantification predominantly rely on SUV thresholding (57, 58); however, manual input requirements and inherent variability limit widespread clinical adoption. AI-driven approaches offer opportunities to enhance clinical workflows through automated quantification of biomarkers, notably TMTV. Fully automated TMTV quantification using whole-body $^{18}$F-FDG PET/CT has been extensively validated across various cancer types (59-62) including prostate cancer (PCa) with prostate-specific membrane antigen (PSMA) PET/CT (63, 64), and neuroendocrine tumors (NET) using $^{68}$Ga-DOTATATE (65, 66), $^{64}$Cu-DOTATATE (67) and $^{18}$F-FDG PET/CT (68). Despite its proven clinical utility, clinical adoption of TMTV remains limited due to the labor-intensive manual segmentation process (52, 69).

LAFOV PET/CT enables advanced PET image quantification such as multi-parametric PET imaging of tumors and organs across the whole body, using streamlined dynamic PET acquisition protocols (22, 70), unlike the complexity of similar protocols on SAFOV PET systems (71, 72). Thus, the need is emerging for the development of robust AI methods tailored specifically for LAFOV data. Various AI techniques for SAFOV, have been developed for anomaly and tumor detection (5) and tumor segmentation (6). These approaches employ single- (PET) and multi-channel-input (PET-CT) neural networks utilizing supervised (73-75), unsupervised/weakly supervised and semi-supervised (76, 77) learning paradigms. In the following section, we briefly review existing AI-based quantification methods for lymphoma, PCa, and NET.

Segmentation of tumors in some cancers such as lymphoma lesions on $^{18}$F-FDG PET/CT is particularly challenging due to lesion heterogeneity and physiological uptake in organs like the brain, liver, and bladder (78). CNN-based segmentation methods (73, 74, 79) have been explored, though AI-driven segmentation methods often struggle with accurately identifying small lymphoma lesions, particularly in limited-stage disease(78). Cascaded AI architectures, involving sequential 2D/3D or 3D/3D processing, have demonstrated promising results by refining initial segmentations (80, 81). Despite good performance, CNN-based methods face challenges in quantifying prediction uncertainty (82), especially under domain shift when a trained model on specific dataset is supposed to be applied on data from other centers (73).

Segmentation of PCa lesions in PSMA PET/CT also poses specific challenges, particularly for metastatic lesions outside the prostate and local recurrences near the urinary bladder, where accurate lesion identification is difficult due to adjacent physiological uptake. While previous AI-driven approaches mostly concentrated on intra-prostatic tumor segmentation (83), recent efforts have explored metastatic lesion detection and segmentation (75, 77, 84).

Studies have associated NET biomarkers, tumor location, and disease type with total tumor volume (65, 68). To address the labor-intensive nature of manual segmentation, AI-based methods such as 3D U-Net CNNs have demonstrated strong agreement with expert segmentations, particularly for hepatic lesions using $^{68}$Ga-DOTATATE PET/CT (85) and total tumor burden with $^{64}$Cu-DOTATATE (67). Additionally, nnU-Net has shown feasibility for automated NET segmentation using $^{68}$Ga-DOTATATE PET/CT, supporting clinical decision-making (66).

Automated segmentation of normal organs can be performed using PET-based approaches (86) or CT-based techniques like TotalSegmentator (87), with resulting masks transferred to PET. Pre-therapy PSMA PET has shown potential for predicting organ doses, enabling personalized prostate-specific membrane antigen radioligand therapy (PRLT) planning (88). Organ segmentation enables organ-based radiomics in FDG PET for cervical and lung cancers (89, 90), and radiomics from negative PSMA PET/CT may help predict progression in recurrent prostate cancer without further treatment.

*Beyond TMTV: The Role of AI-Driven Radiomics in Quantitative Disease Burden Assessment*

While TMTV remains a well-established biomarker for disease burden, recent advances underscore the value of radiomics and AI-driven extraction methods in offering the comprehensive and nuanced characterization of disease. TMTV may underestimate disease burden in diffuse or low-SUV lesions. Radiomics, especially when enhanced with AI tools, can detect and quantify subtle disease patterns that volumetric metrics miss. This supports more accurate patient stratification, particularly in cases where TMTV is less discriminative.

Chauvie et al. (91) and Hasanabadi et al. (92) highlight that radiomics can capture intratumoral and intertumoral heterogeneity, including shape, texture, and intensity-based features. These characteristics reflect biological aggressiveness, treatment response, and dissemination patterns that are not captured by volumetric measures like TMTV alone. Stefano et al. (93) emphasizes the limitations of relying solely on volumetric metrics and calls attention to radiomics as a tool to extract deeper imaging biomarkers. These include first- and second-order texture features, SUV distribution patterns, and spatial relationships among lesions, providing multi-parametric disease burden assessment.

Besides, radiomics allows for the global assessment of disease burden by aggregating features across all lesions or by constructing whole-body imaging signatures. Chauvie et al. (91) argue that this approach yields a more holistic representation of tumor phenotype, which can improve risk stratification and guide therapy planning.

Deep learning techniques, facilitate automated, robust, and reproducible feature extraction, reducing observer variability and improving generalizability. Hasanabadi et al. (92) demonstrate that AI-enhanced radiomics can be integrated into predictive models for survival, progression, and treatment response, serving as quantitative surrogates for disease burden.

*AI-driven Segmentation and Quantification in LAFOV PET/CT*

Precise delineation of volumes of interest (VOIs) in tumors and normal organs remains essential for reliable quantification in LAFOV PET. The improved field-of-view, sensitivity and temporal resolution of LAFOV PET support integration of whole-body time-activity curves, tumor segmentations and kinetic modeling, reducing labor and cost while enhancing clinical applicability (94-97).

Dynamic whole-body PET imaging quality and the quantitative accuracy of the estimated kinetic parameters depend on the completeness of the graphical analysis model(71, 72) or compartmental kinetic model (22, 98) and the sampling of image-derived input functions (IDIFs) as manual arterial blood sampling is invasive and limiting (99, 100). Routine clinical adoption of quantitative methods requires robust AI-guided parametric image generation methods(98, 101) automated AI-based IDIF estimation, particularly in LAFOV PET (99, 102-104). AI-based segmentation tools, such as TotalSegmentator (87) enable accurate segmentation of vascular structures (e.g., aorta and left ventricle), improving IDIF reliability (105, 106).

*Normative PET/CT dataset*

LAFOV PET enables ultra-low dose imaging, allowing safe inclusion of healthy subjects and supporting the development of normative PET/CT databases (107). These databases are particularly valuable in oncology, aligning with the increasing recognition of cancer as a systemic disease (108, 109). While normative databases are common in neuroimaging, extending them to total-body PET is challenged by lower spatial resolution and variable tracer kinetics. Nevertheless, they enable high-throughput screening of at-risk populations and the creation of tracer-specific references for agents such as PSMA, somatostatin receptor ligands, and FAPI. Constructing normative datasets requires accurate deformable alignment of images, where organ segmentation is critical to ensure alignment precision. This can be achieved using advanced AI-based methods, including PET/CT-guided organ segmentation (86), or CT-less approaches (110).

*Enabling the Next Generation of LAFOV PET Imaging Through Collaborative AI Innovation*

Recent AI advances, supported by public datasets like AutoPET, have enabled generalizable tumor segmentation models in PET imaging (111). Realizing the future potential of LAFOV PET with AI requires both robust data-sharing infrastructure and mechanisms for data model standardization and open-access sharing across institutions of both labeled data and AI models. This includes not only access to large, diverse datasets for training foundational models, but also the ability to share and deploy trained models locally for downstream tasks, and to exchange annotated datasets for continual learning. Efficient and scalable annotation is key to this vision, necessitating the integration of user-friendly interactive tools.

Multi-center collaborations and open-source data and AI model sharing, as seen in radiology, are key to building comprehensive normative databases and advancing personalized care. Advancing foundational AI models for PET will benefit from the diverse datasets enabled by LAFOV PET/CT supported through federated learning frameworks (112).

Integration of advanced AI segmentation models such as MedSAM (113) and 3D promptable segmentation (nnInteractive) (114), alongside clinical text data, enhances accuracy and clinical decision-making. Active learning platforms such as MONAI Label (115) enhance the efficiency of annotation workflows, while strategies like expert-reviewed rehearsal and high-uncertainty retraining mitigate model forgetting (115).

Interactive methods, such as DeepEdit (116), reduce annotation burden, and as AI models evolve, reduced reliance on manual corrections will improve consistency and clinical efficiency (117, 118).

## Conclusion

AI is transforming molecular imaging with PET, facilitating advanced segmentation, and enabling automated, quantitative disease burden assessments. The emergence of LAFOV PET/CT necessitates scalable, robust AI methodologies capable of managing extensive datasets and supporting innovative clinical protocols, including ultra-low dose imaging, dynamic multi-parametric studies, and sophisticated radiomics analyses. Foundational AI models and physician-in-the-loop systems further enhance generalizability, clinical applicability, and adoption. However, to realize the full potential of AI-enhanced LAFOV PET imaging, establishing collaborative frameworks for standardized data sharing, federated learning, and comprehensive normative databases is imperative. These developments promise significant advancements in PET image quantification, ultimately enhancing patient stratification, treatment planning, and clinical outcomes.

## Disclosures

Gary Cook has received research support from Siemens Healthineers. Arman Rahmim is a co-founder of Ascinta Technologies. This research was funded by the Canadian Institutes of Health Research (CIHR) Project Grant (PJT-180251), Canadian Institutes of Health Research (CIHR) Project Grant (PJT-173231). All other co-authors have no relevant conflicts of interest.

# TABLES AND FIGURES

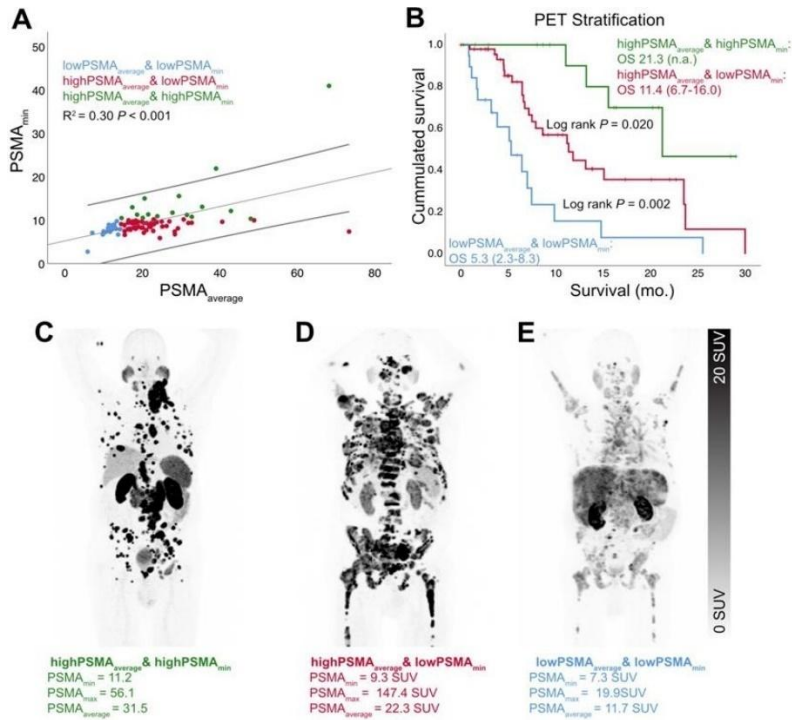

Figure 1: Use of PSMA PET to derive biomarker information. From Ref (43)

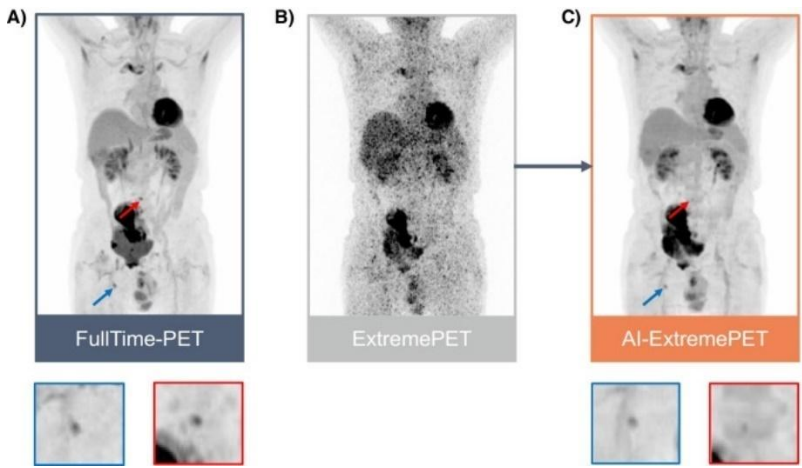

Figure 2: Example imaging quality with enhanced low-dose PET. From Ref (55)